\documentclass[aps,pre,groupedaddress,showpacs,twocolumn]{revtex4}
\usepackage{graphicx}
\usepackage{bm}
\begin{document}
\title{Probing Polyelectrolyte Elasticity Using Radial Distribution Function}

\author{Roya Zandi}
\author{Joseph Rudnick}
\affiliation{Department of Physics, UCLA, Box 951547, Los Angeles,
CA 90095-1547}
\author{Ramin
Golestanian}
\affiliation{Institute for Advanced Studies in Basic Sciences, Zanjan
45195-159, Iran\\
Institute for Studies in Theoretical Physics and
Mathematics, P.O. Box 19395-5531, Tehran, Iran}
\date{\today}

\begin{abstract}

We study the effect of electrostatic interactions on the distribution
function of the end-to-end distance of a single polyelectrolyte chain
in the rodlike limit.  The extent to which the radial distribution
function of a polyelectrolyte is reproduced by that of a wormlike
chain with an adjusted persistence length is investigated.  Strong
evidence is found for a universal scaling formula connecting the
effective persistence length of a polyelectrolyte with its linear
charge density and the Debye screening of its self-interaction.  An
alternative definition of the electrostatic persistence length is
proposed based on matching of the maximum of the distribution with
that of an effective wormlike chain, as opposed to the traditional
matching of the first or the second moments of the distributions.  It
is shown that this definition provides a more accurate probe of the
affinity of the distribution to that of the wormlike chains, as
compared to the traditional definition.  It is also found that the
length of a polyelectrolyte segment can act as a crucial parameter in
determining its elastic properties.

\end{abstract}

\pacs{82.35.Rs, 87.15.La, 36.20.-r, 82.35.Lr}
\maketitle
\section{Introduction and Summary}

The close connection between the elasticity of rodlike cytoskeletal polymers
and the mechanical properties of cells has been extensively documented
\cite{alberts,eichinger,janmey91}.  For example, actin filaments are
known to play a crucial role in the recovery of eukaryotic cell shape
in the face of the stresses imposed by cell movement, growth and
division.  Furthermore, the stiffness of DNA constrains and controls
both its storage in the cell nucleus and access to it by the proteins
and enzymes that are central to its role in biological processes.
However, there is, as yet, no complete, quantitative understanding of
the relationship between the elastic attributes of these polymers and
the influences that control them.  Such an understanding is crucial to
the development of a general description of the functionality of
polymers.

An additional spur to an increased focus on the connection between
first-principle energetics and the elastic characteristics of
polyelectrolytes is the fact that experimental techniques have been
developed that allow for the imaging of single filaments in solution
\cite{Chatenay}.  This means that it is now possible to test models and
theoretical predictions at the level of a single chain polymer.

The key challenge in this area is to produce a theoretical
analysis that connects the electrostatic interactions and local
energetics of a polyelectrolyte with its overall mechanical
properties.  In the case of neutral chain polymers, the wormlike
chain (WLC) model of Kratky and Porod \cite{wlc} provides a
powerful and convenient characterization of flexibility, through
the {\em persistence length}, which is the decay length of the
tangent-tangent correlation function of a stiff, inextensible
polymer chain.  The lack of a similarly inclusive and readily
implemented model for polyelectrolytes has been addressed through
the application of the WLC model to polyelectrolytes, with the
introduction of an effective, ``electrostatic,'' persistence
length \cite{odijk,fixman}.  The shortcomings of this notion,
which replaces the many length scales in a polyelectroyte by a
single effective persistence length, have been commented on
\cite{odijk,joanny,ha}.  Additionally, codifying electrostatic
interactions in terms of single length ignores the possibility of
chain collapse resulting from fluctuation-induced self-attraction
\cite{golestan}.

If a suitably modified WLC model did, indeed, provide a working model
for the calculation of the conformational properties of
polyelectrlytes, then a number of powerful results could be brought to
bear on the study of this system.  Among the more useful of these is
the recently derived end-to-end distribution \cite{degennes} of short
segments of an inextensible polymer \cite{frey}.  The distribution,
which is expected to apply to segments whose backbone length is less
than the persistence length of the polymer---the rodlike limit---was
shown to be accurate both in that limit and somewhat outside it.  The
distribution possesses the virtue that it can, in principle, be
utilized in the analysis of currently feasible experiments on polymers
in the rod-like limit \cite{LeGoff}.  One can, then, determine whether
or not the WLC model accurately reflects the properties of the polymer
segments in question.  If the model has been shown to apply, then the
fit of the distribution to the data directly yields the persistence
length and, hence, the elastic modulus of the polymer.

An additional advantage of the end-to-end distribution function is
that it is a function, rather than a number.  That is, the
end-to-end distribution function contains, at least in principle,
an infinitely greater amount of information than can be encoded in
a single quantity.  The distribution thus represents a more
comprehensive characterization of the elastic and thermal
characteristics of a charged semiflexible chain. In this sense, it
provides one with an opportunity to disentangle the actions of
electrostatic and mechanical interactions as they affect the
conformational properties of a fluctuating polyelectrolyte.

In this article, we calculate the end-to-end distribution of a
charged, inextensible, semiflexible chain.  This distribution is
then utilized to address two central questions relating to the
conformational statistics of a polyelectrolyte.  Those questions
are:
\begin{enumerate}
\item Under what circumstances is the end-to-end distribution of the
PE reproduced by the end-to-end distribution of a WLC with an adjusted
persistence length?

\item Under what circumstances will the effective persistence length
of a PE be as predicted by existing formulas?

\end{enumerate}
As will be shown below, we are able to identify regimes associated
with strong electrostatic coupling, weak screening, or high intrinsic
flexibility of the PE, in which the end-to-end distribution deviates
from that of a WLC. We are also able to identify the regions in which
this approximation is accurate.  The concept of electrostatic
persistence length is examined, and an alternative definition of this
important length scale is suggested.  Our newly-defined electrostatic
persistence length is compared with a well-known and widely-utilized
formula for the persistence length of a stiff, charged rod
\cite{odijk}.  The new electrostatic persistence length is found to
obey a universal crossover scaling law.

The rest of the paper is organized as follows: Section
\ref{sec:model} describes the model that we use to study the
elasticity of stiff PE's. This model is a straighforward extension
of the Kratky-Porod model of a WLC. The energy of this system
consists of the mechanical energy required to induce a local
curvature in the chain and the screened electrostatic repulsion
between the charges on the chain, which are assumed to be
uniformly distributed along it, in the form of a constant linear
charge density.  The distribution function of the end-to-end
distance of this chain is then introduced and given a precise
mathematical form.  An expression for this distribution is
derived, under the assumption that the chain does not undergo too
much of a distortion from it minimum-energy configuration as a
straight line.  That expression, which is based on the eigenvalues
of the Hamiltonian of the PE forms the basis of all the results
reported in this article.  Details of the expansion of the
electrostatic self-interaction about the rodlike configuration are
presented in Appendix \ref{Coulomb}.  Section \ref{sec:eigen} is
devoted to a discussion on the structure of those energy
eigenvalues.  The influence of the electrostatic interaction,
which is most pronounced at on the lower eigenvalues, is
discussed.  Also discussed are the methods and criteria utilized
to guarantee the numerical reliability of the end-to-end
distribution. Section \ref{sec:unscreened} briefly summarizes
exact results that have been obtained for the eigenvalues of the
Hamiltonian consisting entirely of the unscreened Coulomb
interaction. These results are relevant to the conformational
statistics of a chain stiffened entirely by such interactions.  A
more detailed description of the way in which the results were
obtained is relegated to Appendix \ref{app:unscreened}. Section
\ref{sec:comparison} presents the results for the distribution
function in various situations. Particular attention is paid to
the question of the relationship between the distribution function
calculated here and the corresponding end-to-end distribution
function of a neutral rod-like semiflexible polymer.  Section
\ref{sec:elecper} scrutinizes the concept of electrostatic
persistence length.  The point of comparison is a general formula
for the persistence length, introduced by Odijk. We discuss the
ways in which a persistence length can be extracted from the
end-to-end distribution, and we justify the method of choice in
this investigation, which is based on the location of the maximum
of the end-to-end distribution.  We find that our results are
consistent with Odijk's formula when the distribution we calculate
can be made to collapse on the end-to-end distribution of an
uncharged WLC. Conversely, our results do not agree with the Odijk
formula when the distribution is inconsistent with that of a WLC
with adjusted persistence length.  We find that the effective
persistence length of the end-to-end distribution function can be
described by a ``universal'' formula, incorporating a scaling form
and a new, unexpected power law.  This formula holds whether or
not the effective persistence length is described by Odijk's
expression. Section \ref{sec:newtwo} focuses on this behavior of
the effective persistence length.  Section  \ref{sec:moments}
discusses issues related to the moments of the distribution, and
the use of those moments to define a persistence length. Section
\ref{sec:concl} concludes the paper.

\section{Model Elasticity for Polyelectrolytes} \label{sec:model}

Because of the inextensibility of the polyelectrolytes under
consideration \cite{frey,odijk95}, we adopt Kratky and Porod wormlike
chain model \cite{wlc} to describe the bending energy of the chain.
In this model, polymers are represented by a space curve ${\bf r}(s)$
as a function of the arc length parameter $s$.  The total energy of
the chain, which is the sum of the intrinsic elasticity and the
electrostatic energy can be written as

\begin{equation}
\frac{\cal H}{k_{\rm B} T} =\frac{\ell_{p0}}{2} \int_{0}^{L} d s
\left( \frac{d {\bf t}(s)}{ds}\right)^{2} +
\frac{\beta}{2}\int_{0}^{L} d s d s^{\prime} \frac{e^{-\kappa
|{\bf r}(s) - {\bf r}(s^{\prime})|}}{|{\bf r}(s) - {\bf
r}(s^{\prime})|}, \label{energy}
\end{equation}
where ${\bf t}$ is the unit tangent vector.  We do not take into
account the fluctuations in the charges localized to the chain and in
the counterion system that can give rise to attractive interactions
leading to chain collapse \cite{golestan}.  The above equation
contains several different length scales: (1) the average separation
between neighboring charges $b$, (2) the Debye screening length
$\kappa^{-1}$, (3) the Bjerrum length $\ell_{\rm B}=e^{2}/\epsilon
k_{\rm B}T$, the quantity $\epsilon$ being the dielectric constant of
the ion-free solvent, (4) the intrinsic persistence length
$\ell_{p0}$, and (5) the total chain length, $L$.  Two of these length
scales, namely $\ell_{\rm B}$ and $b$, always appear together in the
form of the ratio $\ell_{\rm B}/b^2$, which is denoted by $\beta$ in
Eq.  (\ref{energy}).

We can thus construct three independent dimensionless ratios:
$\beta L$ (a measure of the strength of the electrostatic
interactions), $\kappa L$ (a measure of the degree of screening),
and $\ell_{p0}/L$ (a measure of the intrinsic flexibility).  A
comprehensive exploration of parameter space entails the
examination of the effects associated with a change in the
electrostatic coupling and the degree of screening.  We focus on
inextensible chains that are substantially stiffened by their bending
energy ($\ell_{p0} \sim L$) and for which the bending energy alone is
insufficient to keep the chain nearly rod-like ($\ell_{p0} \ll L$).
The chain is assumed to be sufficiently stiff that excluded volume
does not play a role.

The end-to-end distribution function is defined as follows:
\begin{equation}
{\cal G}({\bf r})=\langle\delta({\bf r}- {\bf R})\rangle,
                 \label{dist0}
\end{equation}
where ${\bf R}={\bf r}(L)-{\bf r}(0)$. The average in Eq.
(\ref{dist0}) is over an ensemble of PE chains.  The function
${\cal G}({\bf r})$ is, then, the probability that a given chain
in the ensemble will have an end-to-end distance equal to ${\bf
r}$.  We make use of the procedure that Wilhelm and Frey have
implemented to calculate the end-to-end distribution function for
inextensible neutral polymers \cite{frey}. In order to calculate
${\cal G}({\bf r})$ in the vicinity of the rodlike limit we
restrict our consideration to regions in which the combination of
intrinsic stiffness and repulsive strength of the Coulomb
interaction keeps the chains in their rodlike limit \cite{note1}.
Using the integral representation of the $\delta$ function, we can
write ${{\cal G} (r)}$ as

\begin{eqnarray}
\lefteqn{{\cal G}({\bf r})  =  {1 \over {\cal Z}}\int \frac{ d^{3}{\bf
\mu}}{(2 \pi)^{3}} \int {\cal D} {\bf t}(s)} \nonumber \\ & & \times
\exp\left\{i {\bf \mu} \cdot {\bf r} - i {\bf \mu} \cdot \int_{0}^{L}
d s \; {\bf t}(s) -\frac{\ell_{p0}}{2} \int_{0}^{L} d s \left(\frac{d
{\bf t}(s)}{ds}\right)^{2}\right. \nonumber \\ && \left. +
\frac{\beta}{2}\int_{0}^{L} d s d
s^{\prime} \; \frac{e^{-\kappa |{\bf r}(s) - {\bf
r}(s^{\prime})|}}{|{\bf r}(s) - {\bf r}(s^{\prime})|} \right\},
\label{dist3}
\end{eqnarray}

\noindent in which ${\cal Z}=\int {\cal D} {\bf t}(s) \;
\exp\left(-{\cal H}/{k_{\rm B} T}\right)$.  We then parameterize
the unit tangent field as
\begin{equation}
{\bf t}(s)={(a_{x}(s),a_{y}(s),1) \over
\sqrt{1+a_{x}^{2}(s)+a_{y}^{2}(s)}},\label{t-a}
\end{equation}

\noindent in which the constraint of inextensibility is manifestly satisfied.

Making use of the relation ${\bf r}(s) -{\bf
r}(s^{\prime})=\int_{s}^{s^{\prime}}d u \;{\bf t}(u)$, we expand the
screened Coulomb interaction about the rodlike configuration to
quadratic order in ${\bf a}(s)=(a_{x}(s),a_{y}(s))$ (see Appendix
\ref{Coulomb} for details).  We then use the series expansion ${\bf
a}(s)=\sqrt{2}\sum_{n=0}^\infty {\bf A}_{n} \cos \left(\frac{n \pi
s}{L}\right)$ as appropriate for the open-end boundary condition, and
assume for simplicity that ${\bf r}$ is, on average, oriented along
the $z$-axis so that ${\bf A}_0={1 \over \sqrt{2}}\int_0^L d s
\;{\bf a}(s)=0$.  Substituting Eq.  (\ref{1distance}) in Eq.
(\ref{dist3}), we obtain
\begin{eqnarray}
\lefteqn{{\cal G}({\bf r}) = {1 \over {\cal Z}}\int \frac{ d^{3}{\bf
\mu}}{(2 \pi)^{3}} \int \prod_{n=1}^{\infty} d^2 A_{n}} \nonumber \\
&& \times \exp \left\{i {\bf \mu}\cdot {\bf r} - i \mu_{z}L +
\frac{i\mu_{z} L}{2}\sum_{n=1}^{\infty}{\bf A}_{n}^{2} \right.
\nonumber \\ && \left.  -\frac{\ell_{p0}}{2L}\sum_{n=1}^{\infty}(n
\pi)^{2} {\bf }A_{n}^{2}-\frac{\beta L}{2}\sum_{n,m=1}^{\infty}{\bf
A}_{n} \cdot {\bf A}_m E_{nm}\right\}, \label{dist7}
\end{eqnarray}

\noindent where

\begin{widetext}

\begin{eqnarray}
E_{nm}&=&{L\over \pi^2 n m}\int_{-\infty}^{\infty} {d k_z  \over 2
\pi} \left(k_z^2+\kappa^2\right)
\ln\left[\frac{(\pi/b)^2+\kappa^2+k_z^2}{\kappa^2+k_z^2}\right] \nonumber \\
&\times& \left\{\cos \left[{\pi \over 2}(n-m)\right]
\left[\frac{\sin\left({k_z L \over 2}\right)\sin\left({k_z L \over
2}-{\pi (n-m) \over 2 }\right)}{k_z \left[k_z-{(n-m) \pi/L
}\right]} -\frac{\sin\left({k_z L \over 2}-{\pi n \over 2}\right)
\sin\left({k_z L \over 2}-{\pi
m \over 2}\right)}{(k_z-{n \pi/L}) (k_z-{m \pi/L})}\right] \right. \nonumber \\
&&\left.-\cos \left[{\pi \over 2}(n+m)\right]
\left[\frac{\sin\left({k_z L \over 2}\right)\sin\left({k_z L \over
2}-{\pi (n+m) \over 2}\right)}{k_z [k_z-(n+m) \pi/L]}
-\frac{\sin\left({k_z L \over 2}+{\pi n \over 2}\right)
\sin\left({k_z L \over 2}-{\pi m \over 2}\right)}{(k_z+n \pi/L)
(k_z-m \pi/L)}\right]\right\} \nonumber \\
&+&{1 \over L}\int_{-\infty}^{\infty} {d k_z  \over 2 \pi}
\left(k_z^2+\kappa^2\right)
\ln\left[\frac{(\pi/b)^2+\kappa^2+k_z^2}{\kappa^2+k_z^2}\right]
\left\{\frac{d^2}{d k_z^2}\left[\frac{\sin^2\left({k_z L \over 2
}\right)}{k_z^2}\right]\right\},\label{Enm}
\end{eqnarray}

\end{widetext}
are the elements of the electrostatic energy matrix in the
cosine basis set, with $b$, the average separation between neighboring
charges, serving as the short distance cutoff.

Performing the functional integral over ${\bf A}_{n}$, we then
obtain

\begin{equation}
{\cal G}({\bf r})={1 \over L} \delta(r_x) \delta(r_y)
\int_{-\infty}^{\infty} {d\omega \over 2 \pi} \; e^{i\omega
(1-r_z/L)} \frac{\det\left[{\bf T}\right]}{\det\left[{\bf
T}+i\omega {\bf I}\right]},
                 \label{integ0}
\end{equation}
where ${\bf T}=(n \pi)^{2}(\ell_{p0}/L) {\bf I} + \beta L {\bf
E}$, ${\bf I}$ is the identity matrix, and $\omega=-{\bf \mu}_{z}
L$. Note that the distribution function in Eq. (\ref{integ0}) is
manifestly normalized to unity, i.e. $\int d^3 {\bf r} \;{\cal
G}({\bf r})=1$. We can then integrate over the orientational
degrees of freedom to find the radial distribution function

\begin{equation}
G(r)=r^2 \int_{0}^{\pi} \sin \theta d \theta \int_{0}^{2 \pi} d
\phi \; {\cal G}({\bf r}),
\end{equation}
that is subject to the normalization $\int_{0}^{\infty} d r \;
G(r)=1$. Using ${\bf r}=(r \sin \theta \cos \phi, r \sin \theta
\sin \phi, r \cos \theta)$, we obtain
\begin{equation}
G(r)={1 \over L} \int_{-\infty}^{\infty} {d\omega \over 2 \pi} \;
\left[e^{i\omega (1-r/L)}+e^{i\omega (1+r/L)}\right]
\frac{\det\left[{\bf T}\right]}{\det\left[{\bf T}+i\omega {\bf
I}\right]},
                 \label{integ0.5}
\end{equation}
which can be rewritten as
\begin{eqnarray}
G(r)&=&{1 \over L} \int_{-\infty}^{\infty} {d\omega \over 2 \pi}
\; \left[e^{i\omega (1-r/L)}+e^{i\omega (1+r/L)}\right] \nonumber
\\ && \times \prod_{n=1}^{\infty} {\left(\lambda_{n} \over
\lambda_{n}+i\omega\right)},
                 \label{integ1}
\end{eqnarray}
in terms of the eigenvalues, $\lambda_{n}$, of the matrix ${\bf T}$.
Contour integration then yields
\begin{equation}
G(r)=\sum_{n=1}^{\infty} \lambda_{n} f(n)
\left[e^{-\lambda_{n} (1-r/L)} \theta(L-r)+e^{-\lambda_{n}
(1+r/L)}\right],
          \label{dist8}
\end{equation}
with
\begin{equation}
f(n)=\prod_{i \neq n}\left({\lambda_{i} \over
\lambda_{n}-\lambda_{i}}\right).
          \label{fn}
\end{equation}

In the above expressions, the second term in brackets contributes
negligibly in the rodlike limit, because even the smallest eigenvalues
are considerably larger than unity in this regime.  Neglecting this
term thus yields

\begin{equation}
G(r)={\cal N} \theta(L-r) \sum_{n=1}^{\infty} \lambda_{n}
f(n) \; e^{-\lambda_{n} (1-r/L)},
          \label{dist9}
\end{equation}
where the normalization prefactor ${\cal N}$ is used to compensate for
the small error caused by neglecting the second contribution.  This
prefactor can be found by enforcing normalization on $G(r)$.

\section{Energy Eigenvalues }
\label{sec:eigen}

\subsection{General Results}

To obtain the radial distribution function we must determine the
spectrum of the effective Hamiltonian ${\bf T}$.  Since this
effective Hamiltonian incorporates the energy cost when the PE
deviates from a rodlike configuration, the spectrum of its
eigenvalues $\lambda_n$ should be positive and a monotonically
increasing function of the index $n$. Note that it is the lowest
eigenvalues that exert the dominant influence on the radial
distribution function.

\begin{figure}[tbp]
\includegraphics[height=2in]{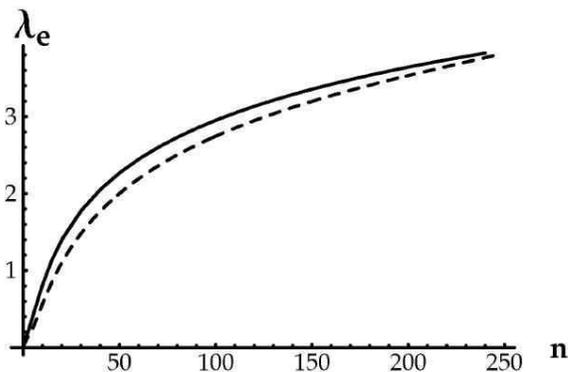}
\caption{Eigenvalues of electrostatic energy matrix calculated from Eq.
(\ref{Enm}), at $L/b=1000$ and $\kappa L=10$ (dashed curve), compared to the
eigenvalues calculated in the infinite length limit in Eq.
(\ref{EnmInf}) (solid curve).}
\label{fig:lfit}
\end{figure}

Figure \ref{fig:lfit} shows the eigenenergies of the electrostatic
interaction ${\bf E}$ at $\kappa L =10$.  To emphasize the importance
of the finite length $L$ of the PE, we calculate the matrix $E_{nm}$
[using Eq.  (\ref{Enm})] in the limit $L \to \infty$, and find
\begin{eqnarray}
E_{nm}(L \to \infty)& = & {1 \over 2} \delta_{nm} \left\{\left[{(n
\pi)^2+(\kappa L)^2 \over (n \pi)^2}\right] \right. \nonumber \\
&& \left.  \ln\left[(n \pi)^2+(\kappa L)^2 \over (\kappa
L)^2\right]-1\right\}.\label{EnmInf}
\end{eqnarray}\noindent
This limit can be achieved in Eq.  (\ref{EnmInf}) by using $\lim_{L
\to \infty} \sin\left[L \left({k_z \over 2 }-{n \pi \over 2
L}\right)\right]/\left({k_z \over 2 }-{n \pi \over 2 L}\right)=2 \pi
\delta(k_z-n \pi/L)$.  Figure \ref{fig:lfit} also displays the
eigenvalues of the diagonal energy matrix in Eq.  (\ref{EnmInf}).  A
comparison of the two curves clearly highlights the importance of end
effects for PE chains of finite length.

\begin{figure}[tbp]
\includegraphics[height=2in]{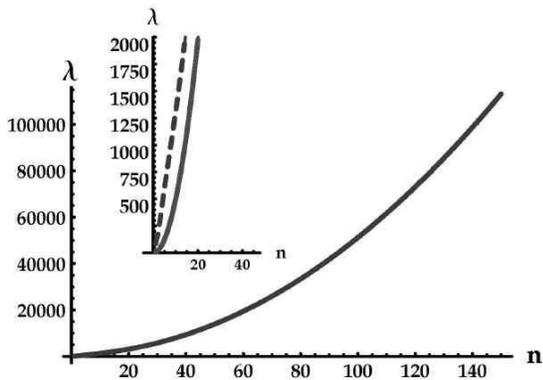}
\caption{Eigenvalues of an elastic charged rod with
$\ell_{p0}/L=0.5$, $\kappa L=0$, and $\beta L=500$.  The bending
eigenenergies are distinguishable from total ones only for lower
values of $n$.  Comparison of the first 20 eigenvalues of bending
(solid line) and total (dashed line) eigenvalues could be seen in
the inset.} \label{fig:eigtot}
\end{figure}

The electrostatic energy comprises one portion of the total energy of
the semiflexible PE. The other contribution arises from the elastic
modulus, quantified by the ``bare'' persistence length, $\ell_{p0}$
and appearing in the energy as the first term on the right-hand side
of Eq.  (\ref{energy}).  The large eigenvalues of the matrix ${\bf T}$
are dominated by the diagonal terms $(n \pi)^{2}(\ell_{p0}/L)$ in, for
instance, the right-hand side of Eq.  (\ref{dist7}).  Thus, the effect
of electrostatic interaction is more pronounced on the lower
eigenvalues, while it is swamped by semiflexible energetics at short
length scales.  In Fig.  \ref{fig:eigtot}, the eigenvalues of the
total Hamiltonian are compared with the bending eigenvalues.  Only the
first few eigenvalues are distinguishable from each other.  The value
of $n$ for which the electrostatic effects can be neglected depends on
the strength of electrostatic interaction $\beta L$ and the intrinsic
persistent length.

The general behavior of the larger eigenvalues associated with
``pure'' electrostatic energy, which grow logarithmically with index,
$n$, is fundamentally different from those resulting to ``pure''
elastic energy, which grow quadratically with index.  In the case of a
{neutral} polymer with an intrinsic persistence length that is
comparable with the total length of polymer, the series in Eq.
(\ref{dist9}) converges very rapidly, as noted in Ref.  \cite{frey}.
When the chain is charged, and especially when the persistence length
is due predominantly to electrostatic effects, more terms in the
series Eq.  (\ref{dist9}) must be preserved in order to obtain a
stable answer for $G(r)$.  In our calculation, the matrix ${\bf T}$ is
truncated at a size much greater than the number of terms needed to
obtain an accurate answer for the series in Eq.  (\ref{dist9}).  There
are two reasons for this.  First, the greater the dimension of the
truncated matrix, the more accurate are the values of lower
eigenvalues which participate in the sum.  Second, retaining more
terms in the product on the right-hand side of Eq.  (\ref{fn}) helps
us achieve a higher accuracy in our result for the function $f(n)$.
In practice, we increased the dimension of the matrices until the
effect (on the end-to-end distribution function) of a further increase
in the size of ${\bf T}$ by a factor of four was less than a part in a
thousand.

The truncation of the matrices above is also related to the short
distance cutoff $b$.  The dimensions of these matrices should not be
larger than $L/b$.  This is because matrices corresponding to a cosine
basis set for which the wavenumber is greater than $1/b$ are
influenced by features at length scales shorter than $b$, and such
features are inconsistent with the inherent coarse-graining in our
energy expression.  We determine the truncation of matrices by looking
for convergence of the numerical procedure, as noted above.  We then
check to see that the size, $N$, of the basis set satisfies $N \le
L/b$.  At no point in our calculations was this inequality violated.

As $r \rightarrow 1$, the dimension of the matrices must be very
high in order to obtain reliable answers.  It is possible to
calculate the integral in Eq. (\ref{integ1}) numerically without
performing contour integration.  This method is very
time-consuming but it is more reliable because this way we reduce
significantly the roundoff error caused by the sum in Eq.
(\ref{dist9}).  We used this approach for very large values of
$\beta L$.

\subsection{Exact Analytical Eigenfunctions and Eigenvalues of the
Unscreened Coulomb Interaction Energy} \label{sec:unscreened}

Although we are not able to produce exact, analytical solutions to
the eigenvalue equations of the quadratic weight in Eq.
(\ref{Enm}), for arbitrary values of the inverse screening length,
$\kappa$, it has proven possible to obtain the eigenfunctions and
eigenvalues of the unscreened version of the energy operator.  The
results of the analytical investigation provide a check for our
numerics and provide insight into the qualitative behavior of
eigenfunctions and eigenvalues of the energy operators controlling the
conformational statistics of rod-like PE's.

As it turns out, the investigations of the properties of the operator
in this special case is most conveniently carried out in ``real
space''---that is, with arc-length, $s$, as the independent variable.
In this section, we will summarize the methods utilized and the
results obtained in this investigation.  The details of the
calculation are presented in Appendix \ref{app:unscreened}.  The key
result is that the eigenfunctions of the unscreened Hamiltonian have
the form
\begin{equation}
\xi_{k}(s) = s^{-1/2} \sin \left(k \ln \frac{s}{b} + \phi(k)
\right) \label{neweigsa}
\end{equation}
and that the eigenvalues are given by
\begin{equation}
\lambda(k)=\Lambda \left( -\frac{1}{2} + ik \right)
\label{lamfuncta}
\end{equation}
where the function $\Lambda(p)$ is defined in Eq.  (\ref{eig1}).  When
the dimensionless wave-vector-like quantity, $k$, is large, the
eigenvalues are essentially linear in $\ln k$.  This is to be
expected, given the form of the unscreened Coulomb interaction, and
is, in fact, consistent with our numerical results.

\begin{figure}[htb]
\includegraphics[height=2in]{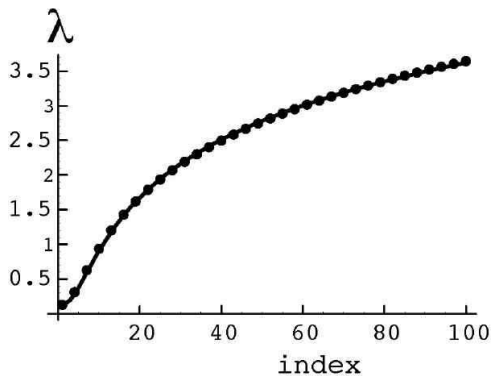}
\caption{Comparison of the eigenvalues $\lambda(k)$ as given by
Eq. (\ref{lamfunct}) with the numerical eigenvalues of the
unscreened Coulomb energy matrix as given by Eq. (\ref{Enm}). As
indicated in the figure, the horizontal axis is the index of the
eigenvalue.  The formula (\ref{lamfunct}) is the solid curve,
passing through the filled circles, which indicate the location of
the eigenvalues of the discrete version of the operator in Eq.
(\ref{Enm}). To facilitate the visual comparison, the numerical
eigenvalues have been ``thinned out,'' in that every third
numerical eigenvalue is shown.} \label{fig:compareeigs}
\end{figure}

\begin{figure}[htb]
\includegraphics[height=2in]{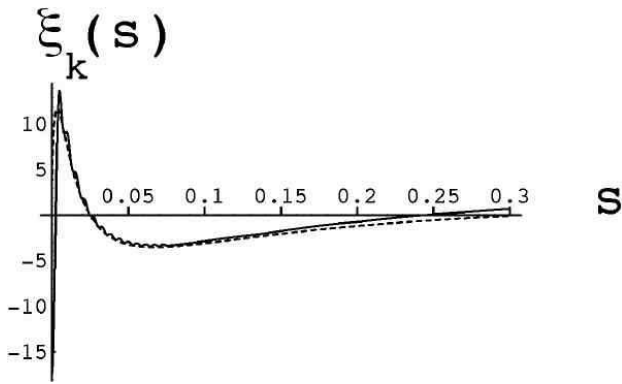}
\caption{Comparison of the numerically determined eigenvector of the
discrete version of the Coulomb energy kernel associated with the
sixth-lowest eigenvalue (solid curve) and of the eigenfunction
generated by the formula (\ref{neweigsa}) (dashed curve).  The value
of the parameter $k$ in that formula is adjusted to optimize the fit
between the two curves, with the use of a least-squares procedure.  The
actual value of $k$ used here is 1.3.  The oscillations in the solid
curve are the result of the cutoff in the cosine basis set.}
\label{fig:eigenvector}
\end{figure}

We can now attempt a comparison between the expressions for the
eigenfunctions in Eq. (\ref{neweigsa}), and eigenvalues in Eq.
(\ref{lamfuncta}), and the numerically generated eigenfunctions
and eigenvalues of the unscreened energy operator whose matrix
elements in the cosine basis are given in Eq. (\ref{Enm}). The
eigenvalues are compared in Fig. \ref{fig:compareeigs}, and a
comparison for a low-lying eigenvector is displayed in Fig.
\ref{fig:eigenvector}.  As indicated by the figures, the numerical
correspondence is excellent.  The quality of the comparison serves
as a validation of the numerical calculations carried out here.

It is possible to assess the importance of a small term representing
the influence of ``pure'' bending energy, by treating that term as a
perturbation of the overall energy of the polyelectrolyte chain. At
first order in perturbation theory, a straightforward calculation
leads to the following result for the ratio of the pure bending energy
contribution to the eigenvalue to the zeroth order energy resulting
from an unscreened Coulomb interaction
\begin{equation}
\frac{E_{\rm bending}}{E_{\rm Coulomb} } = \frac{l_{p0}\left[ ( 1 +
4k^{2}) /4\right] \left[ k^{2}/(1+k^{2}) \right]}{2l_{B} \lambda(k)
\ln(L/b)} \label{ratio1}
\end{equation}
Note that in the limit of an infinitely long polyelectrolyte
chain, the unscreened Coulomb interaction always dominates.  The
divergence in the denominator as $L \rightarrow \infty$ is,
however, quite slow. As practical matter, the bending energy due
to the intrinsic cost of introducing curvature into a
semi-flexible chain will eventually overcome the Coulomb energy as
a contribution to the conformation energy of the chain.  In light
of this fact, one can obtain results for the end-to-end
distribution of an unscreened chain in the presence of a bending
energy.

In this regard, it should be noted that in the absence of a bending
energy, the expression (\ref{integ1}) for the end-to-end distribution
of an unscreened polyelectrolyte chain is formally divergent. This is
because of the very slow increase in the eigenvalues (\ref{lamfuncta}) as
a function of $k$. A full investigation of the conformational
statistics of the unscreened polyelectrolyte chain has not yet been
carried out, and will in all probability yield surprises.

\section{Radial Distribution Function of
Polyelectrolytes}\label{sec:comparison}

\begin{figure}[tbp]
\includegraphics[height=2in]{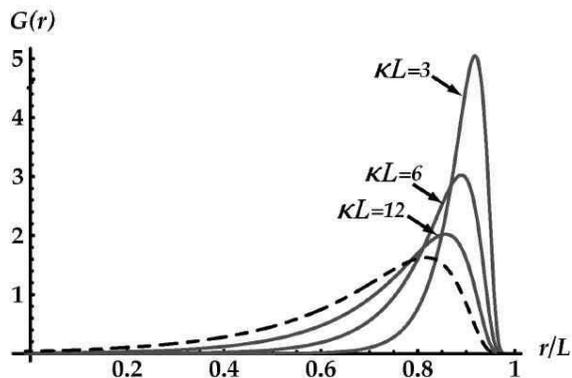}
\caption{Comparison of distribution function of polyelectrolytes
for $\kappa L=12$, $\kappa L=6$ and $\kappa L=3$, respectively
from left to right.} \label{fig:screened}
\end{figure}

Using the eigenvalues of the effective Hamiltonian for the rodlike
PE, we can calculate the radial distribution function from Eq.
(\ref{dist9}). Figure \ref{fig:screened} illustrates the effect of
the screened electrostatic interaction on the distribution
function of a neutral chain (dashed line), for three different
values of $\kappa L =12$, $\kappa L =6$ and $\kappa L =3$.  $\beta
L=100$ for the three PE distributions. As manifested in the
figure, the distribution function of the polyelectrolyte has
shifted toward larger extensions and is peaked more sharply around
the maximum compared to the neutral chain.  Upon decreasing the
screening length, the monomer-monomer repulsion becomes more
strongly screened and the distribution function of the end-to-end
distance peaks towards shorter extensions.

To construct a quantitative measure for the elasticity of
polyelectrolytes, we investigate to what degree and under what
conditions the radial distribution function of a polyelectrolyte
can be reproduced by the distribution of a neutral chain with an
adjusted persistence length.  To this end, we attempt to collapse
the distribution function of a polyelectrolyte onto the
distribution function of a wormlike chain with an effective
$\ell_{p}$, and observe that the answer to the above question
depends on the values of $\ell_{p0}/L$, $\kappa L$ and $\beta L$.

We investigate the effects of overall charging of the chain as
quantified by the combination $\beta L$, of the screening length,
equal to $\kappa^{-1}$, of the length of the chain, $L$ and of the
chain's intrinsic stiffness, described in terms of the bare
persistence length, $\ell_{p0}$.  We are able to identify regimes
in which the conformational statistics of a rod-like PE is
effectively indistinguishable from that of a neutral WLC, and we
also find that there are regions in parameter space in which the
radial distribution of a charged semiflexible rod-like chain
cannot be reproduced by the corresponding distribution of a
neutral chain.

Our conclusions are summarized in the four subsections below.
Briefly, we find that sufficiently strong charging, a sufficiently
long screening length, sufficiently short overall chain length or
sufficiently weak intrinsic stiffness can give rise to deviations
between the conformational statistics of a PE and that of a WLC.

\subsection{Effect of Electrostatic Charging}\label{subsec:charging}

\begin{figure}[tbp]
\includegraphics[height=2in]{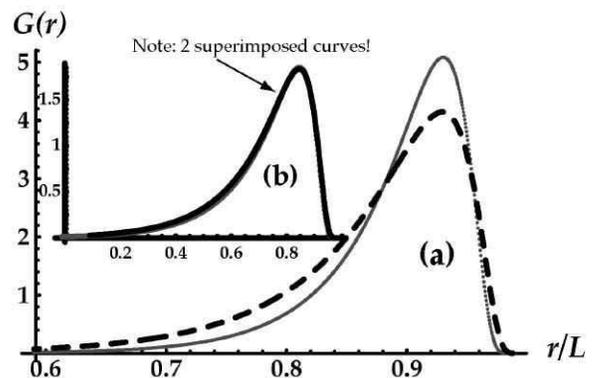} \caption{(a) The
distribution for a polyelectrolyte with $\ell_{p0}=0.5$, $\beta
L=600$, and $\kappa L=10$, compared to that of a wormlike chain
with $\ell_{p}=1.3$.  Inset: (b) The distribution for a
polyelectrolyte with $\ell_{p0}=0.5$, $\beta L=60$, and $\kappa
L=10$, compared to that of a wormlike chain with $\ell_{p}=0.6$
(two superimposed curves).}
      \label{fig:charging}
\end{figure}

There is a substantial regime in the parameter space in which the
distribution of a polyelectrolyte matches exactly that of a wormlike
chain with an adjusted persistence length, particularly when the
electrostatic interaction plays a perturbative role in the chain
energetics.  There are also regimes in which the electrostatic
interaction gives rise to a substantial portion of the stiffening
energy, and in which the radial distribution of the PE is the same as
that of a neutral WLC. For a sufficiently strong electrostatic
interaction, however, the polyelectrolyte distribution begins to
deviate from the wormlike chain form.  This is illustrated in Fig.
\ref{fig:charging} where the polyelectrolyte distribution is compared
with that of the best effective wormlike chain description for the two
cases of low and high charging.  It is clear that at strong enough
charging, the electrostatic energy establishes the conformational
statistics.

\subsection{Effect of Electrostatic Screening}\label{subsec:screening}

\begin{figure}[tbp]
\includegraphics[height=2in]{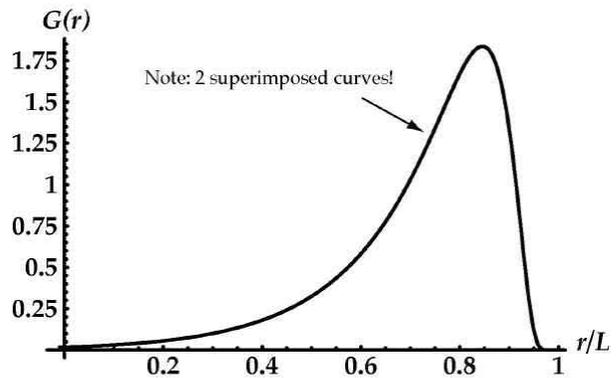}
\caption{The distribution for a polyelectrolyte with $\ell_{p0}=0.5$,
$\beta L=600$, and $\kappa L=50$, compared to that of a wormlike chain
with $\ell_{p}=0.56$ (two superimposed curves).}
      \label{fig:screening}
\end{figure}

The degree of screening of the electrostatic interaction similarly
affects the form of the end-to-end distribution.  In the
high-ionic-strength regime it is possible to obtain a satisfactory
match of the polyelectrolyte distribution with that of an effective
wormlike chain.  However, as screening decreases, the polyelectrolyte
distribution deviates significantly from the wormlike chain form.
Comparison of Figs.  \ref{fig:charging}(a) and \ref{fig:screening}
illustrates the difference between high and low screening for a given
intrinsic stiffness and charging.  We observe that as we increase
screening, the two distribution collapse on top of each other.
Our calculations verify that under physiological conditions ($\kappa=1
\ {\rm nm}^{-1}$), the distribution functions for rod-like DNA
segments ($L \raisebox{-0.03in}{$\stackrel{<}{\sim} $} 100 \ {\rm
nm}$) as well as those of stiffer actin filaments also collapse onto
the end-to-end distribution for neutral chains with an effective
persistence length given by Eq.  (\ref{eqodijk}).

\subsection{Effect of Finite Length}\label{subsec:length}

\begin{figure}[tbp]
\includegraphics[height=2in]{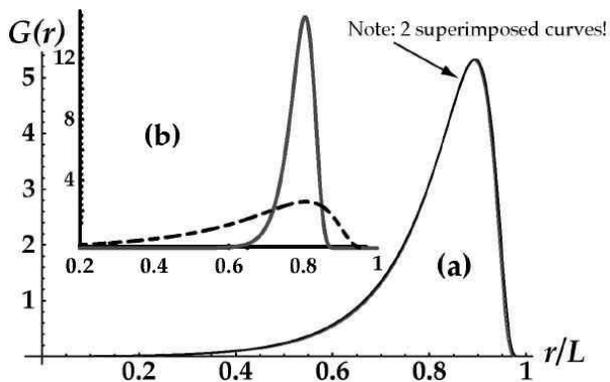} \caption{(a) The distribution
for a polyelectrolyte with $\ell_{p0}/L=0.0001$, $\beta L=36000$,
and $\kappa L=100$, compared to that of a wormlike chain with
$\ell_{p}=0.87$ (two superimposed curves). Inset: (b) The
distribution for a polyelectrolyte with $\ell_{p0}/L=0.01$, $\beta
L=360$ and $\kappa L=1$, compared to that of a wormlike chain with
$\ell_{p}=0.565$. The difference between the two sets of
parameters is only in the length of the chain.} \label{fig:length}
\end{figure}

In most of the studies of the elasticity of polyelectrolytes that
have been carried out to this point, the finite length of the
polyelectrolyte and the corresponding end-effects have been
assumed to be unimportant.  We have re-examined this assumption by
considering two sets of parameters that correspond to identical
values for $\ell_{p0}$, $\beta$, and $\kappa$, and different
values for the contour length $L$ of the polyelectrolytes.  As can
be seen in Fig. \ref{fig:length}, while the distribution for
longer chains can be satisfactorily collapsed onto that of a
wormlike chain, the situation is completely different for shorter
chains.  This highlights the importance of end effects in the
elasticity of polyelectrolytes.

It is important to note that a rescaling of the backbone length of the
PE is not an acceptable stratagem for improving the agreement between
the PE radial distribution and that of a WLC. This is because the
backbone length is essentially fixed by the rod-like chain condition.
The shortening and thickening that is associated with intermediate
blob-like structures \cite{blobs} will not occur.  In fact, we have
observed that a reduction in the effective value of $L$ actually
degrades the quality of the correspondence between the conformational
statistics of a PE and those of the corresponding WLC. In general, our
calculations indicate that the difference between the radial
distribution of the PE and the WLC is partly due to the influence of
end effects.  A complete investigation of these effects is described
in Ref.  \cite{shape}.

\subsection{Effect of Intrinsic Stiffness}\label{subsec:stiffness}

Fig.  \ref{fig:length}(b) indicates that for $\kappa L=1$ and $\beta
L=360$ when the intrinsic persistence length of a chain is small
($\ell_{p0} \ll L =0.01$), the radial distribution function of a
polyelectrolyte does not exhibit near-perfect matching with that of an
uncharged wormlike chain .  In this almost unscreened case, matching
is achieved, when the chain becomes sufficiently stiff.  If
$\ell_{p0}/L$ is increased to 35, while $\kappa L$ remains equal to
one and $\beta L=360$, the radial distribution of a PE will be
indistinguishable from the corresponding distribution of a neutral WLC
with a persistence length $\ell_{p}= 39L$.  The fact that such a large
intrinsic persistence length is required to achieve this kind of
matching of the two distribution highlights the importance of
electrostatic interactions in this case.

\subsection{Applicability of the WLC Model: A General
Diagram}\label{subsec:gen-diagram}

\begin{figure}[tbp]
\includegraphics[height=2in]{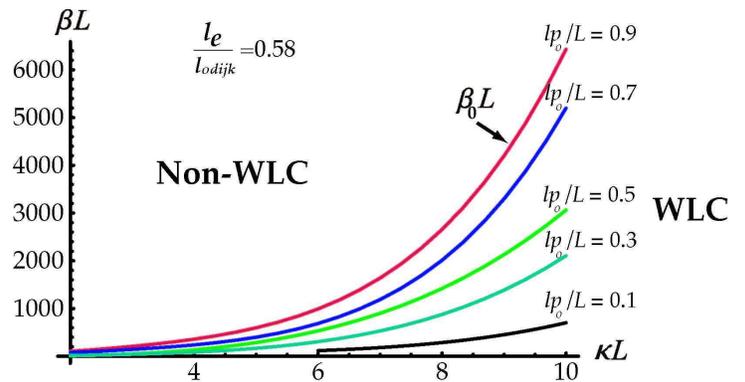}
\caption{The diagram delineating the regimes in which the
statistics of a rod-like PE is the same as that of a WLC with a
suitably modified persistence length from the regimes in which the
statistics of the two models are fundamentally different. The
curves separating the two regimes correspond to $\ell_e/\ell_{\rm
Odijk}=0.58$ for various values of $\ell_{p0}/L$.  These lines
also indicate the values of $\beta_{0} L$ in Eq. (\ref{ours}) at
different values of $\kappa L$ and $\ell_{p0}/L$.  No curves were
drawn outside of the range of rodlike behavior.  For this reason, the
curves for $\ell_{p0}/L=0.1$ and $\ell_{p0}/L=0.3$ are truncated.}
\label{fig:phase}
\end{figure}

The effects discussed above can be summarized in a general diagram
in the three dimensional parameter space. This diagram, shown in
Fig. \ref{fig:phase}, illustrates the conditions under which a WLC
model provides an accurate description of the conformational
statistics of a short, rod-like PE. The lines shown indicate the
locations of the crossover regions that separate the regime in
which one can think of a short segment of a PE as a WLC with a
modified persistence length from the regime in which the
statistics of a PE is fundamentally distinct from that of a WLC.

As can be seen in Fig. \ref{fig:phase}, for a given value of the
screening parameter, charging up the PE will cause a crossover
from WLC into non-WLC behavior. This crossover is, however,
hampered by increased intrinsic stiffness.

\section{Electrostatic Persistence Length}\label{sec:elecper}

In this Section, we investigate the concept of the electrostatic
persistence length.  As in the previous section, we will find that
there are regimes in which Odijk's formula [Eq. (\ref{eqodijk}]
below) for the effective persistence length of a PE \cite{odijk}
is accurate, and that there are regimes in which it does not hold.
As it turns out, there is a strong correlation between the
accuracy of this formula and the correspondence between PE and
neutral WLC conformational statistics.  When the radial
distribution function of a PE is well-reproduced by that of a
neutral WLC, the persistence length of the corresponding WLC is
accurately predicted by Odijk's formula. Conversely, when the two
distributions do not collapse on one another, the persistence
length of the ``best-fit'' WLC is not predicted by that formula.

We have seen in Sec. \ref{sec:comparison} that one can collapse
the distribution function of a polyelectrolyte onto the
distribution function of a wormlike chain with an adjusted
persistence length whenever the Coulomb interaction is no more
than a perturbation to the mechanical stiffness of the chain, or
when Debye screening is sufficiently strong.  In these regimes,
the persistence length of the neutral chain follows Odijk's
prediction, in that, $\ell_{p}=\ell_{e}+\ell_{p0}$, where
$\ell_{p}$ is the effective persistence length of the charged
chain, and the electrostatic persistence length $\ell_{e}$ is
given as \cite{odijk}

\begin{eqnarray}
\ell_{\rm Odijk}& = & \frac{\beta L^2}{12}\left[e^{-\kappa
L}\left(\frac{1}{\kappa L}+\frac{5}{(\kappa L)^{2}}+\frac{8}{(\kappa
L)^{3}}\right) \right.  \nonumber \\ && \left.  +\frac{3}{(\kappa
L)^{2}}-\frac{8}{(\kappa L)^{3}}\right], \label{eqodijk}
\end{eqnarray}
which reduces to $\ell_{\rm OSF}\equiv \beta/4 \kappa^2$ for large
$\kappa L$ \cite{odijk,fixman}.  For instance, at $\beta L= 600$ and
$\kappa L=50$, if we add $\ell_{\rm OSF}=0.06$ to $\ell_{p0}=0.5$, we
find $\ell_{p}=0.56$.  The distribution function of this
polyelectrolyte collapses perfectly onto that of a wormlike chain with
$\ell_{p}=0.56$.  It is important to note that the expression for
$\ell_{\rm Odijk}$ was derived under the assumption that $\ell_{p0}
\sim L$, the contour length of the chain, $L$, is of the order of its
intrinsic persistence length, $\ell_{p0}$ and that the electrostatic
interaction has the limited effect of ``perturbing'' the wormlike
chain shape of the chain \cite{odijk}.  Our results confirm the
validity of OSF formula in the regime in which it is expected to be
correct ($\kappa L \gg 1$).

For long chains, substantial charging is required in order to enforce
the rodlike limit.  It is important to note that despite the
substantial charging of the chain, the ratio of the length of the PE
to the screening length, $\kappa L$, must be sufficiently large.  This
requirement is essential in order to minimize the strong influence of
the end effects on conformational statistics of charged chains.

The important characteristic of this regime is that $\ell_{p0} \ll
\ell_{e}$ and thus electrostatic energy no longer plays a perturbative
role.  However, we find that in this regime Odijk's formula works
perfectly well as well, provided the screening is strong enough.  For
example, at $\ell_{p0}/L=0.0001$, $\beta L=36000$, and $\kappa L=100$,
we find $\ell_{e}/L=0.87$ which is a near-perfect match with
$\ell_{\rm Odijk}$ (see Fig.  \ref{fig:length}).  The accuracy of
Odijk's formula when $\ell_{e} \gg \ell_{p0}$ is not at all obvious,
as OSF was derived in the regime where electrostatic interaction plays
a perturbative role.

We emphasize that the reason for the high quality of the match with
OSF in this regime is different from the reason for the corresponding
result obtained by Khokhlov and Khachaturian \cite{khokhlov} for
weakly charged flexible chains.  In our case, the chain is stiff in
all length scales; the possibility of renormalizing the length
and/or charge is thus excluded in our formulation.

\begin{figure}[tbp]
\includegraphics[height=2in]{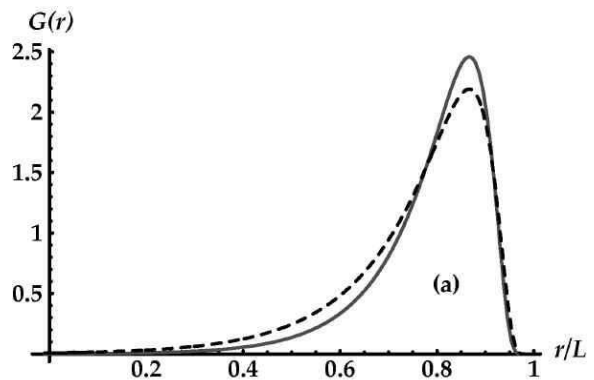}
\caption{The
intrinsic persistence length of the charged chain is
$\ell_{p0}/L=0.5$, $\beta L=45$ and $\kappa L=2$. The persistence
length of the wormlike chain is $\ell_{p}/L=0.767$.}
\label{fig:k2b45}
\end{figure}

As noted in Sec.  \ref{sec:comparison}, increasing the electrostatic
coupling or screening length, one encounters a regime in which it is
no longer possible to obtain near-perfect collapse of the
polyelectrolyte distribution onto that of an effective wormlike chain.
Figure \ref{fig:k2b45}, displays the polyelectrolyte end-to-end
distribution (solid curve) along with the modified wormlike chain
distribution (dashed curve) in a case in which it is possible to
obtain a good, but not perfect fit.  The fit was obtained by matching
the location of the maxima of the two distributions, and the
electrostatic persistence length attributed to the polyelectrolyte
distribution is that of the wormlike chain associated with the dashed
curve.  The PE distribution shown in the figure matches the result of
computer simulation of a rodlike PE \cite{roland}.  In this case the
ratio of electrostatic persistence length extracted from the
distribution function to Odijk's persistence length is,
$\ell_e/\ell_{\rm Odijk}=0.58$.  As noted at the beginning of this
section, the more the two distributions are different from each other,
the more the effective persistence length of the polyelectrolyte
deviates from the Odijk's prediction.  For example, The persistence
length of the wormlike chain shown in Figure \ref{fig:length} (b) is
$\ell_{p}=0.565$, while the effective persistence length of the
polyelectrolyte according to Odijk is $\ell_{p0}+\ell_{\rm
Odijk}=4.52$.  The two distributions differ significantly and the
ratio is $\ell_e/\ell_{\rm Odijk}=0.12$.

Our general observation is that the ratio of $\ell_{e}$ extracted
from the effective wormlike chain distribution to Odijk's
persistence length correlates with the quality of the fit of an
effective wormlike chain end-to-end distribution to that of a
polyelectrolyte. When this ratio is equal to one, the
polyelectrolyte can be completely described in terms of a wormlike
chain. As this ratio decreases, the deviation becomes more
pronounced. Figure \ref{fig:phase} displays a diagram, which
delineates the quality of the fit.  The line that is used in the
Figure to separate the two regimes corresponds to
$\ell_e/\ell_{\rm Odijk}=0.58$. As indicated in this Figure (and
described in Sec. \ref{subsec:gen-diagram}), for fixed $\kappa L$,
when $\beta L$ is below a certain value the polyelectrolyte
behaves like a wormlike chain, while for larger $\beta L$ the two
distributions start to differ significantly. This crossover scale
is sensitive to the intrinsic flexibility of the PE, as shown in
Fig. \ref{fig:phase}.

\begin{figure}
\includegraphics[height=2in]{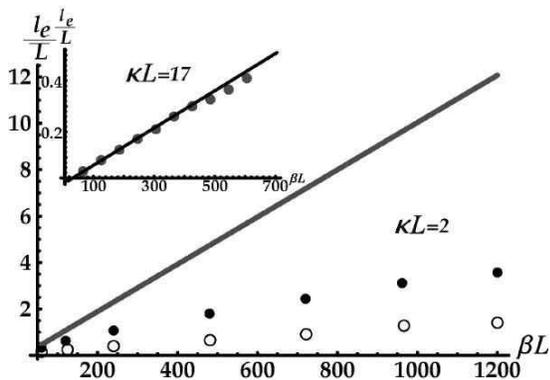}
\caption{Comparison of our results for the electrostatic persistence
length with Odijk's finite size formula (solid line) at
$\kappa L=2$ with $\ell_{p0}/L=0.5$ (filled circles) and
$\ell_{p0}/L=0.01$ (hollow circles).  The inset is for $\kappa L=17$
and $\ell_{p0}/L=0.5$.} \label{fig:comp}
\end{figure}

There is a substantial difference between the electrostatic
persistence length of a rodlike polyelectrolyte and Odijk's
prediction as we increase the electrostatics strength $\beta L$.
As illustrated in Fig.~\ref{fig:comp}, the deviation of
$\ell_{e}/L$ from $\ell_{\rm Odijk}/L$, with increasing $\beta L$
is more pronounced at lower values of $\kappa L$.  At $\kappa
L$=17, this deviation becomes evident when $\beta L\sim 1000$.
However, at $\kappa L=2$, the deviation is evident already when
$\beta L\sim 30$.

A review of the current literature on elasticity of
polyelectrolytes reveals that there is no simple theory for
computing $\ell_{e}$ with arbitrary intrinsic stiffness.  It is
generally believed that as long as a polyelectrolyte chain is
stiff, the dependence of $\ell_{e}$ on $\kappa$ is correctly given
by the OSF formula. However, our results indicate that in certain
regimes the electrostatic persistence length, $\ell_{e}$, depends
on $\ell_{p0}$, and is smaller than is predicted by the OSF
formula. Figure ~\ref{fig:comp} shows that the electrostatic
persistence length of a rodlike polyelectrolyte, $\ell_{e}/L$,
depends on $\ell_{p0}/L$. This dependence is not present in
Odijk's formula.

Our general observation of the curves of $\ell_{e}$ versus $\beta
L$ is that they asymptote to a power law of the form
$\ell_{e}\propto (\beta L) ^{x}$ where the exponent $x(\kappa L)$
is less than $1$. This behavior has been predicted by Barrat and
Joanny with an exponent $x=1/2$ in the limit $\beta L \to \infty$
\cite{joanny}.  We find, however, that the asymptotic limit is
approached much more slowly than it is predicted by the
Barrat-Joanny crossover formula.  The section below describes our
attempt to systematize the relationship of the effective
persistence length to Odijk's prediction \cite{itamar}.

\section{Universal Behavior}\label{sec:newtwo}

\begin{figure}[tbp]
\includegraphics[height=2in]{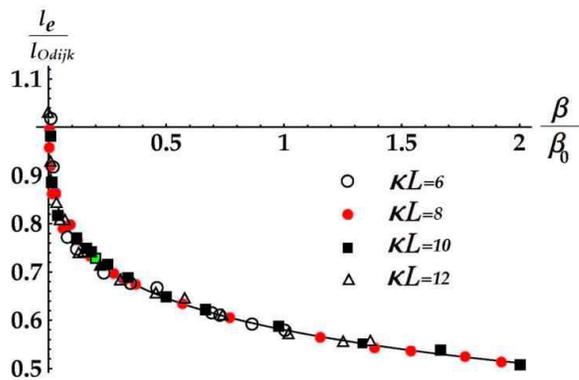}
\caption{Comparison of our data, suitably rescaled, to the
expression in Eq. (\ref{ours})
for the form of the correction to Odijk's formula for
the electrostatic persistence length.
The points correspond to our rescaled plots of $\ell_{e}/L$ versus
$\beta L$ at $\kappa L=6,8,10,12$.}
\label{fig:rescale}
\end{figure}

In search of a possible universal pattern which might relate
$\ell_{p0}/L$, $\beta L$ and $\kappa L$ to $\ell_{e}$, we have
rescaled our graphs for different values of the parameters.  To this
end, we choose a value of $\beta$ called $\beta_{0}$ such that
$\ell_{e}/\ell_{\rm Odijk}=0.58$, and rescale the parameter $\beta$
with it, where the value of 0.58 is completely arbitrary.  We have
followed this procedure to rescale our graphs for different values of
$\kappa L$, keeping $\ell_{p0}/L$ constant.  Figure \ref{fig:rescale}
contains these rescaled graphs for $\kappa L=6,8,10,12$, which
collapse satisfactorily on top of each other.  It is important to note
that when we choose any other ratio of $\ell_{e}/\ell_{\rm Odijk}$ to
rescale our graphs, we obtain the same universal behavior among our
curves.  The solid line in Fig.  \ref{fig:rescale} corresponds to the
following crossover or interpolation formula for $\ell_{e}$

\begin{equation}
\ell_{e}=\frac{\ell_{\rm Odijk}}{1+c(\beta/\beta_{0})^{x}},
                  \label{ours}
\end{equation}
plotted with the exponent $x=0.4$.  In the above equation,
$c=0.724$ is constant which ensures the ratio of
$\ell_{e}/\ell_{\rm Odijk}$ for a given value of $\beta_{0}$.

\begin{figure}[tbp]
\includegraphics[height=1.7in]{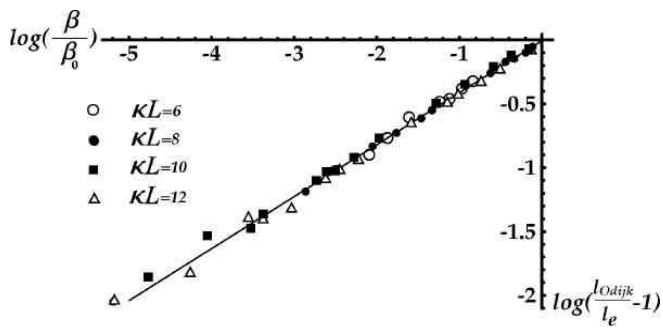} \caption{Demonstration of
the universal scaling behavior for $\ell_{e}$. The points
correspond to our rescaled plots of $\ell_{e}/L$ versus $\beta L$
at $\kappa L=6,8,10,12$.  The equation of straight line is
$y=0.006+0.401x$.} \label{fig:logscale}
\end{figure}

As exemplified in Fig.~\ref{fig:rescale}, this formula provides a
remarkable fit to our data.  To calculate the exponent $x$, we have
attempted to fit our curves of $\log(\ell_{\rm Odijk} / \ell_{e}-1)$
versus $\log(\beta/\beta_{0})$ to a straight line.  As shown in Fig.
\ref{fig:logscale}, all our data are fitted to a straight line with
slope approximately equal to 0.4.  Similar curves have been generated
for other, smaller, values of $\ell_{p0}/L$.  The behavior of these
other curves is the same as the one plotted in Fig.~\ref{fig:rescale}.

The quantity $\beta_{0}$ is an increasing function of $\ell_{p0}$
and $\kappa$.  We have been able to obtain the value of
$\beta_{0}$ from our data.  The crossover discussed in the
previous section occurs when $\beta=\beta_{0}$ which corresponds
to $\ell_{e} \simeq 0.58 \ell_{\rm Odijk} $ in Eq. (\ref{ours}).
Hence, the curves in Fig.~\ref{fig:phase} indicate the values of
$\beta_{0} L$ for different values of $\kappa L$ and
$\ell_{p0}/L$. The dependence of $\beta_0$ on the parameters
$\ell_{p0}$, $\kappa$, and $L$ seems to be in general quite
complicated.  For $\ell_{p0} \kappa >1$, we seem to find that
$\beta_{0} \sim L^2 \kappa ^{4} \ell_{p0} \; g(\kappa L)$, where
$g$ is a polynomial function. For $\ell_{p0} \kappa < 1$, however,
the leading dependence on $\kappa$ seems to be a much stronger
power law.  We are still investigating the dependence of $\beta_0$
on the parameters in all the different regimes \cite{rudnick}.

\section{First and Second Moments of the Distribution}  \label{sec:moments}

Much theoretical work on electrostatic persistence length, is
based on the calculation of $\langle r \rangle$ or $\langle r^{2}
\rangle$ of a polyelectrolyte. Odijk uses the second moment to
test the validity of OSF when $\ell_{e}>\ell_{p0}$ \cite{odijk}.
It is thus crucial to learn to what extent the conformational
statistics of polyelectrolytes are similar to those of wormlike
chains when they share the same $\langle r \rangle$ or $\langle
r^{2} \rangle$.  It is not obvious that the matching of first or
second moments of the end-to-end distribution will yield the
matching of the higher moments of the distribution as well.

We extract an ``effective'' persistence length by matching the first
or second moments of the end-to-end distance of a polyelectrolyte
distribution with that of an effective wormlike chain in order to
compare our results with other existing theories.

\begin{figure}[tbp]
\includegraphics[height=2in]{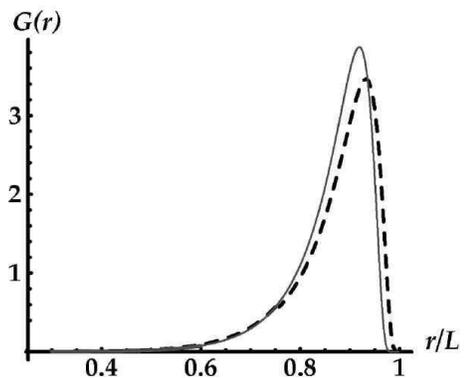}
\caption{The taller distribution (solid line) belongs to a
polyelectrolyte at $\kappa L=17$, $\ell_{p0}=0.5$ and $\beta
L=1000$.  The shorter distribution (dashed line) is adjusted so
that its second moment is the same as that of the polyelectrolyte
distribution.} \label{fig:sc17r2}
\end{figure}

Figure \ref{fig:sc17r2} contains the end-to-end distribution function
for a screened polyelectrolyte with $\kappa L=17$, $\ell_{p0}/L=0.5$
and $\beta L=1000$ along with the distribution function of an
uncharged wormlike chain with $\ell_{p}/L=1.22$.  The shorter
distribution is adjusted so that its second moment is the same as that
of the polyelectrolyte distribution.  It is noteworthy that the
persistence length of the uncharged wormlike chain agrees perfectly
well with Odijk's prediction, while the two distribution functions are
distinguishable from each other.  In general, we observe that the
maximum of the wormlike chain distribution is shifted toward larger
extension when we match $\langle r^{2} \rangle$ of the two
distributions as exemplified in the Figure.

\begin{figure}[tbp]
\includegraphics[height=2in]{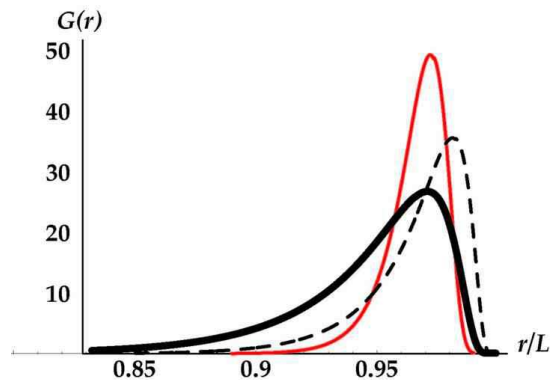} \caption{The thin solid line
is the plot of $G(r)$ for a polyelectrolyte with $\beta L=600$,
$\kappa L=0$, and $\ell_{p0}/L=0.5$.  The thick solid line with
$\ell_{p0}/L=3.1$ belongs to a WLC distribution for which its maximum
is matched to that of the PE distribution.  The dashed plot is for a
neutral chain with $\ell_{p0}/L=4.86$.  The dashed distribution and
polyelectrolyte have the same first moment.}
\label{fig:unscr1}
\end{figure}

Similar results are observed when we match the first moment of
distribution of a polyelectrolyte with that of a wormlike chain
with an adjusted persistence length.  In Fig. \ref{fig:unscr1},
the distribution function of a polyelectrolyte along with two
distribution functions for effective wormlike chains is presented.
The dashed line and polyelectrolyte share the same first moment.
As illustrated in the Figure, the two distributions differ
significantly.

\begin{figure}[tbp]
\includegraphics[height=2in]{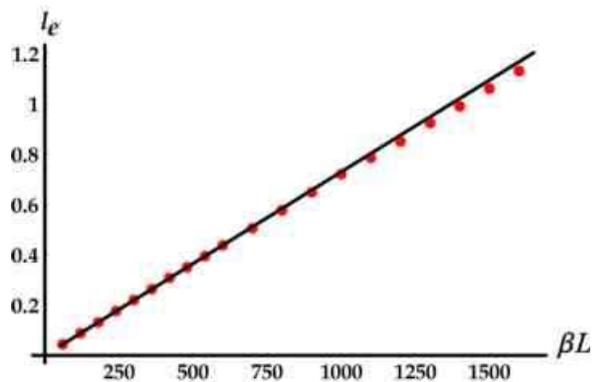}
\caption{Comparison of our results for the electrostatic
persistence length with Odjik's finite size formula [
Eq.(\ref{eqodijk})] (solid line) at $\kappa L=17$.  Our numerical
results are based on matching of the second moment of
distributions.} \label{fig:comp17r2}
\end{figure}

We have compared the electrostatic persistence length given in Eq.
(\ref{eqodijk}) with our results obtained by matching $\langle r^{2}
\rangle$ of a PE distribution to that of a WLC distribution.  Figure
\ref{fig:comp17r2} shows that for $\kappa L=17$, our data begins to
deviate from Odijk's finite size formula when $\beta L \sim 1200$.

It is important to note that when we match the maximum of
distributions to extract an effective persistence length, the
deviation from Odijk's prediction begins at $\beta L \sim 400$ for
$\kappa L=17$ (the inset in Fig.  \ref{fig:comp}).  This is the point
when the two distributions cease to closely resemble each other.
However, if we use the matching of the first or second moment to
calculate the effective persistence length, the two distributions
would clearly be distinguishable from each other well within the
regime in which Odijk's formula for the persistence length remained
accurate.

This points to the fact that replacing a PE chain by a WLC, when they
share the same $\langle r^{2} \rangle$ or $\langle r \rangle$, is not
well-justified in all regimes and that one should use care in the
utilization of the notion of an electrostatic persistence length based
on matching of first or second moments.

\section{Conclusion}    \label{sec:concl}

The difficulty in producing a complete characterization of the
mechanical properties of a charged, semiflexible chain arises from the
existence of a number of length scales in that system.  In particular,
it is far from clear that the notion of a persistence length provides
an adequate description of the mechanical and thermal characteristics
of such a chain.  In light of this, the end-to-end distribution
function provides an attractive alternative.  Given that the
distribution is a function, rather than a single number, it represents
considerably more information about the charged chain.

One important application of this distribution is in the
assessment of the utility of the notion of a persistence length,
in that it is possible to compare the distribution obtained
experimentally, via simulations, or as the result of explicit
calculations, with the corresponding distribution of a neutral
semiflexible chain.  On the basis of explicit calculations, we
have been able to determine the extent to which the end-to-end
charged chain can be collapsed onto the end-to-end distribution of
a neutral worm-like chain with an adjusted persistence length.
Among the regimes in which this correspondence is achieved are
those in which electrostatic effects play a perturbative role.  In
addition to collapse of the two distributions in these regimes, we
also find that the electrostatic persistence length is given by
the formulas of Odijk and OSF. This result is consistent with the
assumptions underlying those formulas.  We also observe collapse
of the distributions and can verify the validity of the Odijk and
OSF results for the effective persistence length in regimes in
which the charging of the chain is sufficiently strong that
electrostatic effects dominate purely mechanical energetics as
long as Debye screening of electrostatic interactions is
sufficiently strong.

In fact, we find that there is a strong correlation between
correspondence of the distributions of charged and uncharged chains
and the validity of Odijk's formula for the effective persistence
length.  When the distributions can be collapsed onto each other, the
formula proves to be accurate, while lack of correspondence of the two
distributions is accompanied by inaccuracy of the Odijk result for the
effective persistence length.  Our results indicate that the
difference between the radial distribution of the PE and the WLC can
be attributed, at least in part, to the influence of end effects.  In
fact, we believe that the behavior of the persistence length is
substantially controlled by end effects.  One way of understanding
this is in terms of Odijk's derivation of the expression
(\ref{eqodijk}) for the electrostatic persistence length \cite{odijk}.
This derivation is based on a calculation of the energy of a bent
segment of a charged rod.  A key assumption in this derivation is that
the segment takes the form of an arc of a circle.  End effects are
readily associated with the difference between the shape of a real
bent rod and the circular arc assumed in Odijk's derivation.  An
exploration of these effects in this context is described in Ref
\cite{shape}.

Another important finding is that an effective persistence length,
obtained by locating the maximum of the distribution, can be described
in terms of a scaling formula, Eq.  (\ref{ours}).  This formula
relates the actual persistence length to the Odijk predictions.  The
formula is ``universal,'' in that it has a general form that is
independent of the parameters utilized, and it incorporates a power
law that does not appear to be anticipated in the Hamiltonian
governing the system, nor does it arise from any simple dimensional
analysis.  At this point, we have no explanation for either the
universal form or the power law.

As noted above, the effect of counterion condensation has been
ignored throughout the above work.  It has been shown that
counterion condensation modifies the bending rigidity of a
semiflexible chain \cite{golestan,kremer,rouzina,haim} and may
result in the collapse of the PE chain \cite{golestan}.  We have
performed a calculation of the distribution function taking into
account the attractive interaction due to counterion fluctuations
and observed the signature of collapse. In general, one is able to
observe the collapse of a charged chain for any strong enough
attractive interaction which is in function of the distance
between monomers.  We are currently investigating these
effects\cite{rudnick}.

\acknowledgments

The authors would like to acknowledge helpful discussions with
W.M. Gelbart, M. Kardar, R.R. Netz I. Borukhov, H. Diamant, K. -K.
Loh, V. Oganesyan, and G. Zocchi. This research was supported by
the National Science Foundation under Grant No. CHE99-88651.

\appendix

\section{Expansion of the Coulomb Interaction}  \label{Coulomb}
In this Appendix, the expansion of the Coulomb self-interaction
for a charged chain that is slightly deformed about the rodlike
configuration is derived. Using Fourier representation of the
screened Coulomb interaction we find
\begin{widetext}
\begin{eqnarray}
\int_0^L d s \int_0^L d s' \frac{e^{-\kappa |{\bf r}(s) - {\bf
r}(s^{\prime})|}}{|{\bf r}(s) - {\bf r}(s^{\prime})|}&=&\int
\frac{d^3 {\bf k}}{(2 \pi)^3} {4 \pi \over k^2+\kappa^2} \int_0^L
d s \int_0^L d s' \; \exp\left\{i {\bf k} \cdot \left[ {\bf r}(s)
- {\bf r}(s^{\prime})\right]\right\} \nonumber \\
&\simeq&\int \frac{d^3 {\bf k}}{(2 \pi)^3} {4 \pi \over
k^2+\kappa^2} \int_0^L d s \int_0^L d s' \; \exp\left\{i {\bf
k}_{\perp} \cdot \int_{s}^{s'} d u \; {\bf a}(u)-{i k_z \over 2}
\int_{s}^{s'} d u \; {\bf a}(u)^2+ i k_z (s'-s)\right\} \nonumber \\
&=&\int \frac{d^3 {\bf k}}{(2 \pi)^3} {4 \pi \over k^2+\kappa^2}
\int_0^L d s \int_0^L d s' \; e^{i k_z (s'-s)} \nonumber \\
&& \;\times\; \left[1-{1 \over 2 }\left({\bf k}_{\perp} \cdot
\int_{s}^{s'} d u \; {\bf a}(u)\right)-{i k_z \over 2}
\int_{s}^{s'} d u \; {\bf a}(u)^2+O(a^3)\right].
       \label{1distance}
\end{eqnarray}
\end{widetext}

\section{General form of the energy associated with distortions of
a rod-like polyelectrolyte stiffened entirely by unscreened Coulomb
interactions}
\label{app:unscreened}

Suppose that the tangent vector ${\bf t}(s)$ is as given by Eq.
(\ref{t-a}). Suppose, also, that the interaction energy of the
rod-like polyelectrolyte is given by
\begin{equation}
{\cal E} = \frac{\beta}{2} \int_{0}^{L}ds \int_{0}^{L }ds^{\prime}
V(\textbf{r}(s) - \textbf{r}(s^{\prime})). \label{en1}
\end{equation}
Making use of Eq. (\ref{t-a}) and the relationship
\begin{equation}
\textbf{r}(s) = \int_{0}^{s}\textbf{t}(s_{1})ds_{1}, \label{rdef}
\end{equation}
and expanding to the resulting expression for the energy ${\cal E}$ to
second order in the deviation, $\textbf{a}(s)$, from a straight line
in th $z$-direction, we find that the energy is given by
\begin{eqnarray}
{\cal E} & = & \frac{\beta}{2} \left\{ \int_{0}^{L} ds \int_{0}^{L
} ds^{\prime} V(s-s^{\prime}) \right. \nonumber \\ && \left.  +
\int_{0}^{L} ds\int_{0}^{L} ds^{\prime} {\cal K}(s,s^{\prime})
\textbf{a}(s) \cdot \textbf{a}(s^{\prime}) \right\}, \label{en2}
\end{eqnarray}
where
\begin{equation}
{\cal K}(s,s^{\prime}) = \delta(s-s^{\prime}) \int_{0}^{L} {\cal
L}(s,s^{\prime}) ds^{\prime} - {\cal L}(s,s^{\prime}),
\label{Kdef}
\end{equation}
and
\begin{equation}
{\cal L}(s, s^{\prime} ) = -\int_{0}^{s_{<}} ds_{a}
\int_{s_{>}}^{L} ds_{b}
\frac{V^{\prime}(s_{b}-s_{a})}{s_{b}-s_{a}}, \label{Ldef}
\end{equation}
and $s_{>(<)}$ is the greater (smaller) of $s$ and $s^{\prime}$. If
the charges on the polyelectrolyte interact via the unscreened Coulomb
potential, then the interactions leading to the kernel ${\cal
L}(s,s^{\prime})$ are straightforward, and we find
\begin{equation}
{\cal L}(s,s^{\prime}) = \frac{1}{2} \left[\frac{1}{s_{>}-s_{<}} +
\frac{1}{L} - \frac{1}{s_{>}} - \frac{1}{L-s_{<}} \right].
\label{Lun1}
\end{equation}
Note that the kernel ${\cal L}(s,s^{\prime})$ is equal to zero
whenever either one of the arguments is equal to $0$ or $L$.

The kernel ${\cal K}$ is neither local nor is it translationally
invariant. However, if we are interested in what happens in the
vicinity of $s=0$, we can simplify ${\cal L}$ and, as a
consequence, ${\cal K}$. When both $s$ and $s^{\prime}$ are much
smaller than $L$, we can replace ${\cal L}$ as given by Eq.
(\ref{Lun1}) by
\begin{equation}
{\cal L}_{n}(s,s^{\prime}) = \frac{1}{2} \left[
\frac{1}{s_{>}-s_{<}} - \frac{1}{s_{>}} \right]. \label{Lun2}
\end{equation}
The kernel ${\cal K}$ that results from this new ${\cal L}$ via Eq.
(\ref{Kdef}) is still non-local and is not translationally invariant.
However, it is possible to obtain, by inspection, its eigenfunctions
and their associated eigenvalues.

We assume that the minimum spacing between adjacent charges, $b$, is
small, and we assume that we can retain those terms that are leading
order in ratios of $b$ to other lengths in the system.  Making use of
these assumptions, we are able to obtain, by inspection, results for
the eigenfunctions and eigenvalues of the energy operator ${\cal K}$.

\begin{figure}[htb]
\includegraphics[height=2in]{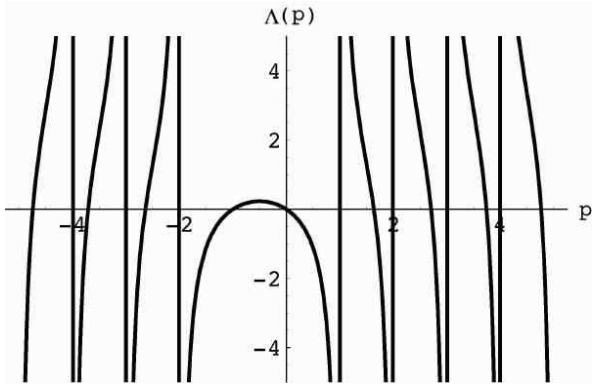}
\caption{ the function $\Lambda(p)$ as a function of real $p$.}
\label{fig:pgraph}
\end{figure}

\begin{figure}[htb]
\includegraphics[height=2in]{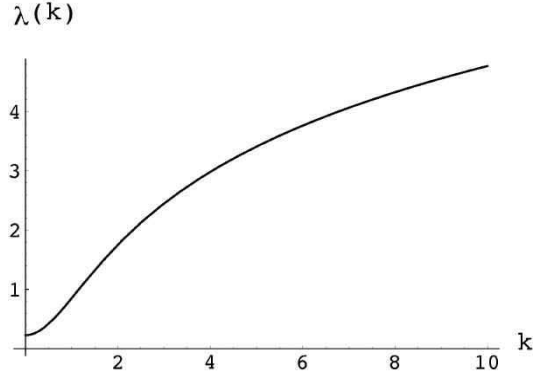}
\caption{The function $\lambda(k) = \Lambda(-1/2+ik)$, for $k>0$.}
\label{fig:lamgraph}
\end{figure}

We start by noting that the convolution of the energy kernel in Eq.
(\ref{Lun2}) with the function $s^{p}$ is equal to
\begin{eqnarray}
s^{p} \left\{2 \ln \frac{s}{a} +p \int_{0}^{1} \alpha^{p-1}\ln
(1-\alpha) d \alpha - \right. \nonumber \\ \left.  p
\int_{1}^{\infty} \alpha^{p-1} \ln (\alpha -1) d \alpha
-\frac{1}{p+1} + \frac{1}{p}\right\}. \label{coeff1}
\end{eqnarray}
The assumption underlying the calculations leading to (\ref{coeff1})
is that the integrand is fundamentally convergent.  That is, we ignore
the possibility that the integrand yields non-integrable divergences
anywhere in the region of integration. We now make use of the
following relations
\begin{equation}
\int_{0}^{1} \alpha^{p-1}\ln (1-\alpha) d \alpha = -\gamma -
\psi(p+1), \label{firstint}
\end{equation}
\begin{equation}
\int_{1}^{\infty} \alpha^{p-1} \ln (\alpha -1) d \alpha = \gamma +
\psi(-p), \label{secondint}
\end{equation}
where $\psi(x)$, the digamma function, is the logarithmic derivative
of the gamma function:
\begin{equation}
\psi(x) = \frac{1}{\Gamma(x)} \frac{d}{dx} \Gamma(x).
\label{psidef}
\end{equation}
This means that the ``eigenvalue'' associated with the eigenfunction
$s^{p}$ is given by
\begin{equation}
\Lambda(p) =2 \gamma + \psi(1+p) + \psi(-p) +\frac{1}{p+1} -
\frac{1}{p} -1. \label{eig1}
\end{equation}
Figure \ref{fig:pgraph} display the function $\Lambda(p)$ as a function
of the real argument $p$.
Note that the function is symmetric about $p=1/2$.  It can be
readily shown that the function $\Lambda(p)$ is real if $p=-1/2
\pm ik$, with $k$ real.  The proper eigenfunctions and eigenvalues
are associated with just such values of $p$.  In fact, we can
choose for eigenfunctions of the operator
\begin{equation}
\xi_{k}(s) = s^{-1/2} \sin \left(k \ln \frac{s}{b} + \phi(k)
\right). \label{neweigs}
\end{equation}
If we then require that the derivative of the eigenfunction
$\xi_{k}(s)$ is zero at the boundary ($s=b$), then
$\phi(k)=\arctan 2k$.  It can be shown that the integrations
leading to the eigenvalue are all convergent, and the eigenvalue
has the form $\Lambda(-1/2 + i k)$. A plot of this eigenvalue as a
function of the parameter $k$ is displayed in Figure
\ref{fig:lamgraph}.
This eigenvalue is an even function of $k$. As it turns out, an
excellent numerical approximation to
\begin{equation}
\lambda(k) = \Lambda(-1/2 + ik), \label{lamfunct}
\end{equation}
is given by
\begin{equation}
\lambda(k) \approx 0.2274112+\ln \left[ 1+ 0.929 k^{2} \right]
\label{eigapprox}
\end{equation}
The spacing of the eigenvalues is determined by the allowed values
of the parameter $k$ in (\ref{lamfunct}).  If we assume that the
eigenfunctions pass through zero at some large value of $s$, or
that they have a zero slope there, then it is straightforward to
show that the lowest eigenvalues will be associated with
equally-spaced $k$'s. Figure \ref{fig:compareeigs} contains a
comparison between the eigenvalues given by Eq.  (\ref{lamfunct})
with equally-spaced $k$'s and the numerically calculated
eigenvalues of the interaction operator with matrix elements in
the cosine basis set as shown in Eq. (\ref{Enm}).

As an additional check on the validity of the results presented
here, we compare the eigenfunctions in Eq. (\ref{neweigs}) to the
eigenvectors of matrix in Eq. (\ref{Enm}).  The eigenvectors are
plotted in real space. We find that for low eigenvalues, the
expressions in Eq. (\ref{neweigs}) provide an excellent match to
the results of numerical calculations.  Figure
\ref{fig:eigenvector} displays a comparison between an
eigenfunction as given by Eq. (\ref{neweigs}) and the
numerically-determined eigenvector associated with the
sixth-lowest eigenvalue of the discrete version of the kernel. The
value of $k$ in the formula for the eigenfunction was adjusted
with the use of a least-squares procedure.  The eigenfunctions
have zero slope at the boundary of the region associated with the
smallest value of $s$.  The quantity $s$ does not range over the
entire interval, from 0 to 1, because the analytic eigenfunctions
generated here are expected to be accurate only in the regime $s
\ll L$, where $L$ has been set equal to one in the case at hand.

\begin{figure}[htb]
\includegraphics[height=2in]{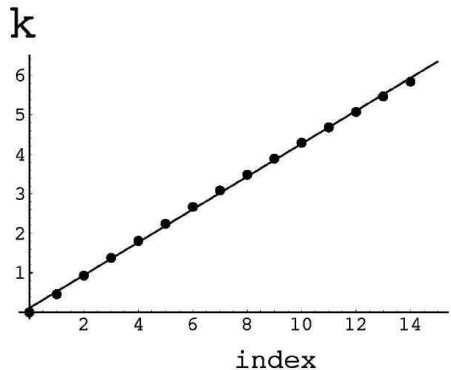}
\caption{The values of the parameter $k$.}
\label{fig:kvals}
\end{figure}

The least-squares fitting of the analytical eigenfunctions to
numerical eigenvectors leads to a set of values for the parameter,
$k$.  Figure \ref{fig:kvals} displays the values of $k$, plotted
against the index of the eigenvalues.  As indicated by the
straight line drawn through them, the $k$'s are approximately
equally-spaced.

\pagebreak

\bibliographystyle{apsrev}
\bibliography{poly}

\begin{thebibliography}{24}
\bibitem[{not()}]{note1}
\bibinfo{note}{Numerical evaluation of $G(r)$ for neutral semiflexible chains
  based on a Monte-Carlo simulation in Ref. \cite{frey} reveals that the
  approximation taken for analytical calculation of $G(r)$ is good when
  $L/\ell_{p}\leq 2$. Thus we restrict our attention to this regime and assume
  that $\ell_{p}/L= 0.5$ is the onset of rodlike limit, where $\ell_{p}$ is an
  extracted effective persistence length for the case of charged chains.}
\end{thebibliography}

\end{document}